\documentclass[conference, 10pt]{IEEEtran}
\IEEEoverridecommandlockouts
\usepackage{cite}
\usepackage{amsmath,amssymb,amsfonts}
\usepackage{algorithmic}
\usepackage{graphicx}
\usepackage{svg}
\usepackage{textcomp}
\usepackage{xcolor}
\usepackage{authblk}
\usepackage{amsmath}
\usepackage[export]{adjustbox}

\usepackage{paralist}

\usepackage{graphicx}

\usepackage{subfig}

\usepackage{fancyhdr}
\def\BibTeX{{\rm B\kern-.05em{\sc i\kern-.025em b}\kern-.08em
    T\kern-.1667em\lower.7ex\hbox{E}\kern-.125emX}}
\begin{document}

\title{Effects of Forward Error Correction on Communications Aware Evasion Attacks\\
\thanks{This material is based upon work supported by the National Science Foundation under Grant Number 1303297. Any opinions, findings, and conclusions or recommendations expressed in this material are those of the author(s) and do not necessarily reflect the views of the National Science Foundation.}
}

\author[1]{Matthew DelVecchio}
\author[2]{Bryse Flowers}
\author[1]{William C. Headley}
\affil[1]{Hume Center for National Security and Technology, Virginia Tech}
\affil[2]{Department of Electrical and Computer Engineering, University of California, San Diego}

\maketitle

\begin{abstract}
Recent work has shown the impact of adversarial machine learning on deep neural networks (DNNs) developed for Radio Frequency Machine Learning (RFML) applications. While these attacks have been shown to be successful in disrupting the performance of an eavesdropper, they fail to fully support the primary goal of successful intended communication. To remedy this, a communications-aware attack framework was recently developed that allows for a more effective balance between the opposing goals of evasion and intended communication through the novel use of a DNN to intelligently create the adversarial communication signal. Given the near ubiquitous usage of forward error correction (FEC) coding in the majority of deployed systems to correct errors that arise, incorporating FEC in this framework is a natural extension of this prior work and will allow for improved performance in more adverse environments. This work therefore provides contributions to the framework through improved loss functions and design considerations to incorporate inherent knowledge of the usage of FEC codes within the transmitted signal. Performance analysis shows that FEC coding improves the communications aware adversarial attack even if no explicit knowledge of the coding scheme is assumed and allows for improved performance over the prior art in balancing the opposing goals of evasion and intended communications.
\end{abstract}

\begin{IEEEkeywords}
radio frequency machine learning, adversarial machine learning, forward error correction, cognitive radios
\end{IEEEkeywords}

\section{Introduction}
Deep learning has been shown to be a transformative technology for providing substantial improvements in modalities such as image recognition, natural language processing, text-to-speech, among many others. Given this fact, in recent years a significant research push has been occurring in applying deep neural networks (DNNs) to wireless communication applications. In particular, the research area of Radio Frequency Machine Learning (RFML) has seen exponential growth. RFML is typically defined as utilizing DNNs in order to solve complex wireless communication problems of interest using raw, or preprocessed, IQ sample data as input. Recent research has shown that RFML has led to improved capabilities over traditional solutions in areas such as spectrum sensing, modulation scheme creation, emitter identification, and automatic modulation classification (AMC) \cite{RN9, RN13, RN14}.

Given this explosion in systems using DNNs, much scrutiny has been directed at the security of these learning-enabled systems, a study which is generally termed as adversarial ML. More specifically, the class of adversarial ML attacks known as evasion attacks have been shown to be particularly potent in emphasizing the inherent vulnerabilities of RFML solutions (see \cite{RN1, RN19, RN29, RN22, RN30}, among others). Put simply, evasion attacks are approaches by which the input data is manipulated with intelligently crafted perturbations in order to lower the confidence of a target DNN's outputs, induce incorrect classifications, or even target those classification decisions at a specific label of choice by the attacker \cite{RN16, papernot, RN18}. 

Traditional evasion attacks typically assume that these intelligently crafted perturbations are created and applied directly at the input to the DNN. However, for most real-world applications, this assumption of direct access to the DNN is impractical. Therefore, for RFML applications, there are two key considerations that must be considered. First, the evasion attack must be resilient to the propagation channel between the signal transmitter and the DNN. Secondly, and perhaps most importantly, is the fact that the perturbations must minimize their impact on the intended receiver. While the previous works have considered the first goal in detail, there has been little work to date on consideration of the second goal. To remedy this, the authors' prior work developed a communications-aware framework that provides a mechanism for balancing the conflicting goals of successful communication and DNN evasion \cite{RN2}.

This work extends the communications-aware evasion attack framework through the consideration of forward error correction (FEC) coding. Due to the usage of FEC coding in the vast majority of real-world communication systems to correct errors that arise due to hardware impairments, channel propagation effects, etc., incorporating FEC in this framework is a natural extension of this prior work and is shown to improve performance in more adverse environments. More specifically, this work shows that the intelligently crafted perturbations inherently learn to utilize the nature of the coding to limit the negative impacts of altering the original signal while having a negligible impact on the attack's performance on the target DNN. In other words, this work shows that this intelligent perturbation can be learned without providing explicit knowledge of the FEC in the perturbation creation process. While the communications-aware framework is discussed herein in the context of AMC, it can be generalized for other RFML applications of interest. 

\begin{figure*}
\centering
    \includegraphics[width=\linewidth]{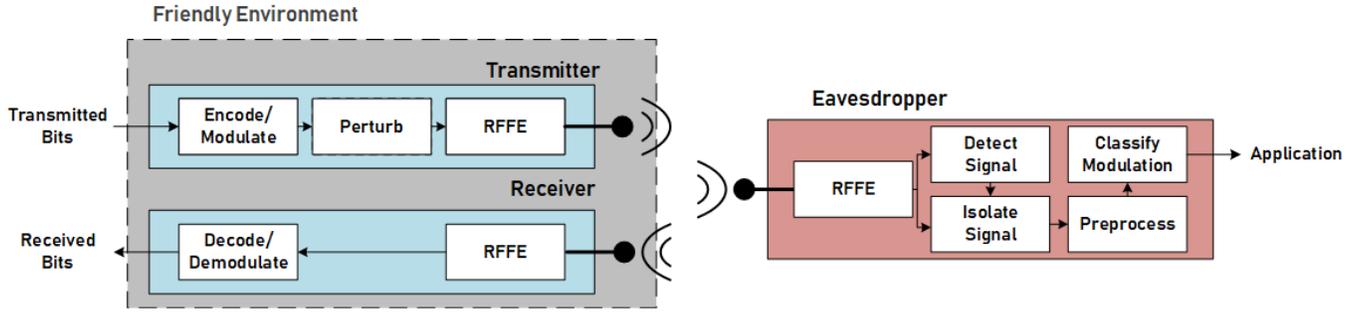}
    \caption{The wireless communications scenario considered within this work in which an intended communications link is being eavesdropped. The "perturb" block of the transmitter utilizes the developed communications aware attack framework to perturb the transmitted signal to evade the eavesdropper.}
    \vspace{-1em}
    \label{fig:background:ThreatModel}
\end{figure*}

This paper is broken down as follows. Section II provides a description of the overall system model and environment assumed in this attack. Section III lays out the framework for the communication-aware attack that utilizes FEC. The results of this work and an analysis of the transmitter's ability to craft smarter perturbations with FEC is shown in Section IV. Finally, this work concludes with a discussion of the key findings of this research as well as directions for future work.

\section{System Model}

Figure 1 depicts the wireless communications scenario for this work which consists of three main components: a transmitter, an intended receiver, and an eavesdropper. 

\subsection{Transmitter}
The transmitter has two competing goals within this scenario: to successfully communicate with a naive receiver, and to evade modulation classification by an AMC-based eavesdropper. Here, the two metrics used to evaluate the success of these goals are, respectively, the bit error rate (BER) at the receiver and the modulation classification accuracy of the eavesdropper. Within the communications aware framework, the transmitter balances between these goals by deploying a form of adversarial machine learning known as an evasion attack. In an evasion attack, the signal ingested by the deep learning algorithm under attack is intelligently perturbed in order to fool the algorithm. In this work, these perturbations are produced by a specially trained network called an Adversarial Mutation Network (AMN). This is represented by the "Perturb" block in Figure 1 and is described in Section III. 

Here, the transmitted data is assumed to be modulated using a linear digital-amplitude phase modulation scheme (ASK, PSK, QAM, etc.) and is pulse shaped using a root-raised cosine filter. Unlike the previous work, the data is assumed to be encoded using an FEC code in order to add redundancy for correcting errors induced by the propagation channel and/or due to the evasion attack's perturbations themselves. In particular, block codes are assumed within this work. Block codes are first assumed in this work as they allow for a more recognizable pattern of redundancy during encoding, given that they encode blocks of bits independently, which the AMN should be well suited to learn without explicit knowledge of the code. As an initial step, learning without explicit information of the coding is desirable so that the communications aware attack framework can easily be executed on different coding types without needing to change the architecture of the AMN, or attack framework, which is useful in modern automatic modulation and coding approaches.


In order to guarantee that the eavesdropper's performance is not impacted by the FEC code itself, the transmitter also uses data whitening after FEC coding to whiten the bits and create a more uniform distribution of bits during transmission. More specifically, this is done in order to guarantee that the eavesdropper is only impacted by the ability of the communications aware framework to create intelligent perturbations signals rather than the inherent difference between FEC-enabled and non FEC-enabled signal structures. This work uses an IBM implementation from \cite{RN27}.

\subsection{Receiver}

Simply put, the role of the receiver is to successfully receive the transmitted data. In this work, the receiver is assumed to be static and unaware of the perturbations being applied at the transmitter. This models a scenario in which the transmitter is adaptable to changes in the environment (such as the presence of an eavesdropper) but the receiver is a legacy system that cannot be easily adapted on the fly. Additionally, for ease of analysis and preliminary performance analysis, it is assumed that the receiver and transmitter are synchronized (e.g. through the use of a header and/or control channel). 

\subsection{Eavesdropper}

The eavesdropper utilizes a state-of-the-art RFML approach to perform modulation classification using the raw IQ data sent by the transmitter. The eavesdropper is assumed to have limited knowledge of the transmitted signal and therefore must detect, isolate, pre-process the raw data prior to modulation classification. This work focuses on disrupting the classification stage of this processing chain. More specifically, in this work it is assumed that the eavesdropper uses a convolutional neural network (CNN), like that presented in \cite{RN9}, to perform modulation classification. The eavesdropper's CNN was trained using synthetic signal data created using LiquidDSP \cite{RN17} with SNRs ranging from 0-20 dB five modulation schemes (described in the following). Like the receiver, the eavesdropper is assumed to have no knowledge of the communications aware framework and therefore does not react to the attack. 

\begin{figure*}
    \includegraphics[width=1.0\linewidth, center]{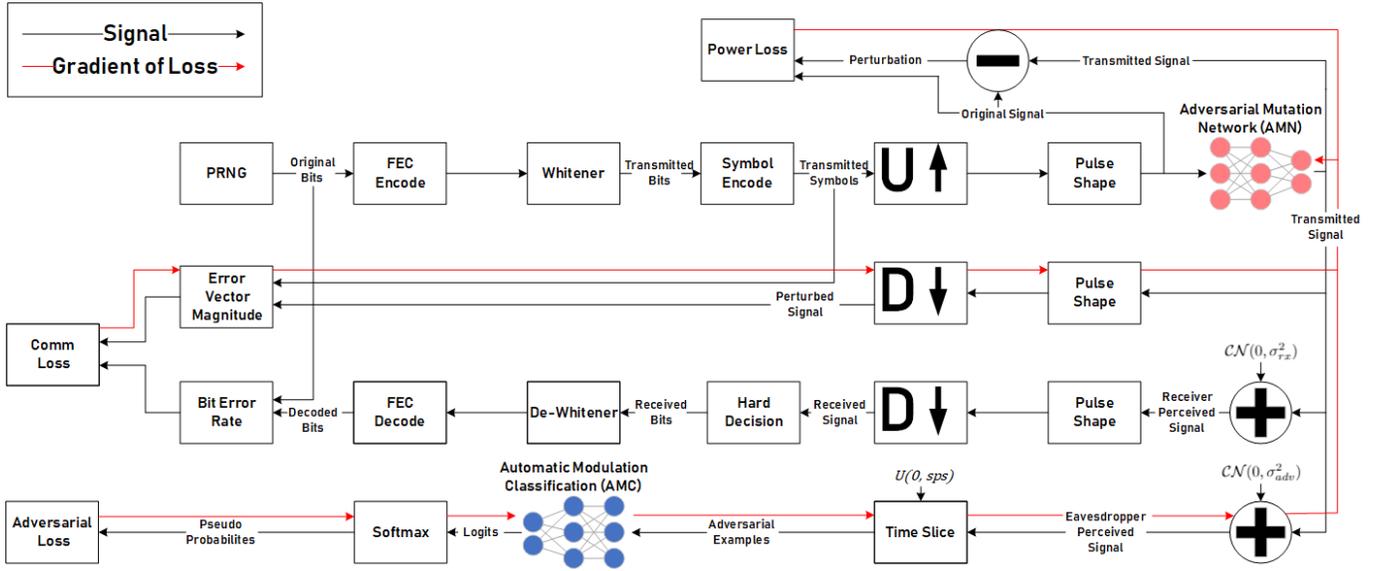}
    \caption{The communications aware attack framework training process. Three loss functions (power loss, communications loss, and adversarial loss) are utilized by the AMN during the training process to intelligently craft the signal perturbations for the given spectral environment.}
    \vspace{-1em}
    \label{fig:background:ThreatModel}
\end{figure*}

\subsection{Data and Environmental Assumptions}

In this work, without loss of generality, the modulation schemes considered are BPSK, QPSK, 8-PSK, 16-QAM, and 64-QAM. For each of these modulation schemes, the communications aware attack framework is trained and its performance is evaluated. The propagation channel between the transmitter and both the receiver and eavesdropper is assumed to be modeled by an AWGN channel. For both training and testing the framework, the SNRs at the receiver and eavesdropper are assumed to be uniformly distributed between 5 and 15 dB. The SNRs and noise realizations between the transmitter and receiver and the transmitter and eavesdropper are assumed to be independent from one another. Additionally, an integer sample time offset is introduced as a channel effect for the eavesdropper in order to assume asynchronous operation with the transmitter. This time offset is assumed to be uniformly distributed between 0 and 8 (the assumed number of samples per symbol). Prior work has shown that time offsets larger than the samples per symbol have little effect on the success of the adversarial attack \cite{RN1}. 

\section{Communication Aware Framework}

This section presents the methodology used to carry out a communications aware attack that utilizes FEC to deceive an eavesdropper while maintaining effective communication between the transmitter and receiver. A description of the nature of the AMN that is used to create the perturbed signal is provided first in this section. Then the custom loss functions used in this work to balance the goals of communication and evasion are discussed. Finally, an overall explanation of the training and testing procedure is presented.

\subsection{Adversarial Mutation Network}

In previous work, various gradient-based approaches have been used to generate perturbations of the original transmitted signal to achieve the goal of classification evasion \cite{RN16, RN24}. While effective, these techniques typically focus on the success of the evasion at the detriment to the communication link between transmitter and receiver. This process involves perturbing a signal using a loss gradient that is back-propagated through the classifier network to create an adversarial signal. This may have to be done multiple times per signal block and must be done separately for every signal block transmitted, creating a very computationally-exhaustive process in a wireless transmitter that it typically resource constrained. Alternatively, networks known as Adversarial Transformation Networks (ATN) make use of a separate neural network to generate the perturbation automatically \cite{RN25}. These networks can utilize custom loss functions in order to balance the adverse effect that perturbations have on communication. Additionally, while the training process of a network requires a large number of computations, once trained it only requires one forward pass of a signal through the network in order to create a perturbed signal to be transmitted, which is much more computationally efficient than solutions relying on gradient-based optimization. 

The work presented in \cite{RN25} provides two examples of such a network: Adversarial Auto-Encoders (AAE) and Perturbation- Adversarial Transformation Networks (P-ATN). While based on similar concepts, the output of these two networks are different. An AAE is provided an input signal and outputs the adversarial signal to be sent over the air. This signal is a learned combination of the original signal plus a perturbation and is represented by the equation
\begin{equation}
    x^{*} = g(\theta, x)
    \label{eq:aae}
\end{equation}
where $g(\cdot)$ denotes the AAE, $\theta$ represents the parameters of the network learned during training, and $x$ is the original signal. A P-ATN is given the same input signal as the AAE but instead outputs a perturbation that is then added to the original signal for transmission. The following equation shows this process.
\begin{equation}
    x^{*} = x+ g(\theta, x)
    \label{eq:p_atn}
\end{equation}
In summary, the AAE crafts the complete signal while the P-ATN creates a perturbation to be added to a signal. 

Previous work in this area used a P-ATN to accomplish the goals of the transmitter \cite{RN2} but this work shifts to the use of an AAE. While P-ATN implementations have simpler convergences (a P-ATN would only need to output 0 in order to transmit a signal optimal for communication while an AAE would need to learn to pass the original signal through the network unchanged), these introduce a different problem. The scaling of the perturbation with respect to the original signal must be done outside of the network. The P-ATN doesn't explicitly learn this scaling process. An AAE inherently learns to balance the power of the perturbation and original signal since it outputs the combined signal and therefore simplifies the process. Scaling is important because if the perturbation was significantly powerful, the transmitter would have an unfair advantage over the eavesdropper. Additionally, most communications are limited in the power that they can transmit. For this reason, this work imposes a power limit on the AMN. The current work will utilize an AAE and refer to this as an Adversarial Mutation Network (AMN) throughout the paper.  The architecture of this network consists of a three convolutional layers with \emph{tanh} activation functions in between these layers. The AMN takes in a single-channel complex input of size [1, 1, 2, N] for a signal with N samples and outputs a signal of the same size and dimensions. 

\subsection{Loss Functions}

The AMN utilizes custom loss functions during training in order to achieve its goal of deceiving the classification abilities of the eavesdropper while simultaneously limiting the BER at the receiver. This communications-aware attack improves the ability of previous evasion attacks that only focused on misclassification. Due to the dueling nature of the transmitter's  focuses of communication and evasion, multiple loss functions must be used that are then balanced. The total loss function is shown below.
\begin{equation}
    \mathcal{L}_{\text{total}} = \alpha \mathcal{L}_{\text{adv}} + \beta \mathcal{L}_{\text{comm}} + \gamma \mathcal{L}_{\text{pwr}}
    \label{eq:total_loss}
\end{equation}

The current work uses three loss sub-functions: adversarial loss, communications loss, and power loss denoted as $\mathcal{L}_{\text{adv}}$, $\mathcal{L}_{\text{comm}}$, and $\mathcal{L}_{\text{pwr}}$ respectively. The adversarial loss seeks to minimize the eavesdropper's ability to successfully classify the signal, the communications loss looks to minimize the BER impact at the receiver, and the power loss seeks to minimize the power of the perturbation, thus keeping the adversarial signal similar to the original signal. These losses are summed together to provide an overall loss metric for the training process. In order to balance the effects of each of these loss terms, balancing constants are set for each of the losses. This helps set the desired tradeoff between communication success and classification evasion. The three constants, $\alpha$ for $\mathcal{L}_{\text{adv}}$, $\beta$ for $\mathcal{L}_{\text{comm}}$, and $\gamma$ for $\mathcal{L}_{\text{pwr}}$, are varied with respect to each other and sum to 1. As $\alpha$ grows, the transmitter becomes more focused on evasion and the resulting signal tends towards noise. As $\beta$ increases, communication improves at the detriment of the evasion ability. As $\gamma$ grows, the signal instead converges to the original, unperturbed signal. 

Each of the separate loss functions are constructed so that they converge to 0 when achieving their desired effect, as is typical of other loss functions used to train DNNs. The optimization technique Adam, which utilizes gradients of the loss, is used during training which therefore requires the loss functions be differentiable \cite{RN28}.

\subsubsection{Adversarial Loss}

Adversarial loss looks to maximize the ability of the AMN to evade classification by the eavesdropper. In this sense, the intent of the adversarial loss metric mirrors that of a loss that would be used in more traditional evasion techniques such as FGSM. The metric used is the confidence the eavesdropper network has that the received signal is the original source modulation, determined by a softmax output. A decrease in this confidence can lead to a successful untargeted attack when the source class is no longer the one determined as most probable by the classifier. 
The loss function used is rooted in log-likelihood and approaches 0 as the confidence decreases but tends toward $\infty$ as the confidence increases. The adversarial loss is defined as 
\begin{equation}
    \mathcal{L}_{\text{adv}} = -log(1-p_{s})
    \label{eq:l_adv}
\end{equation}
where $p_{s}$ represents the confidence of the original source modulation scheme and is obtained as the result of a softmax activation layer at the output of the eavesdropper's AMC network. This work only evaluates untargeted attacks but could easily be used to implement a targeted attack using $-log(p_{t})$. 


\subsubsection{Communication Loss}

The intent of the attack presented in this work is to carry out an attack that evades classification by a malicious eavesdropper while simultaneously allowing for effective communication between a transmitter and receiver. While the adversarial loss metric described above helps accomplish the former, it has a negative impact on the latter. As the AMN has increased success at fooling the eavesdropper, this typically means that the transmitted signal is very different from the original and therefore the communication reliability will degrade. The communication loss is included to guide the AMN to find a balance between adversarial success and communication success.

One way to quantify the reliability of communication is with BER. In the system model used in this work, the bits are retrieved by making a hard decision on the received symbols. Unfortunately, this process is not differentiable which means a gradient can't be calculated for the training process. If the hard decision process was differentiable, this loss could simply be an MSE between the original and received bits or could be defined by a gradient formula as shown in \cite{RN29}. In order to circumvent this issue, the communication loss utilizes two components, BER and error vector magnitude (EVM) of the symbols. The loss is defined as
\begin{equation}
    \mathcal{L}_{\text{comm}} = b_{r} \times EVM(S_{tx}, S_{tx+p})
    \label{eq:l_comm}
\end{equation}
where $b_{r}$ is the BER calculated using the signal received over the noisy channel after the bits have been decoded with FEC. Since the encoding used in this work is done at the bit level instead of encoding the symbols, the error calculations are also performed on the bits rather than on symbols as was done in the prior work. The EVM shown in the loss function represents the distance between the original symbols and the noiseless symbols after the perturbation is added. The EVM is calculated as $|S_{tx} - S_{tx+p}|$. EVM is used in the loss because it is differentiable; thus the BER serves as a magnitude while the EVM serves as a direction for the weights to update.

\subsubsection{Power Loss}

The third and final loss component used in this work is the power loss. The power loss aims to reduce the power of the perturbation compared to that of the original signal. The loss is calculated using an altered version of the signal-to-perturbation ratio (SPR), given as
\begin{equation}
    \mathcal{L}_{\text{pwr}} = \frac{1}{E_{s}/E_{p}} = \frac{E_{p}}{E_{s}}
    \label{eq:l_pwr}
\end{equation}
where $E_{p}$ is the energy of the perturbation and $E_{s}$ is the energy of the signal. The ratio is switched so that it decreases as the perturbation energy decreases, mirroring the behavior needed for the loss function. Additionally, the above equation is done on a linear scale rather than logarithmic so that the values are between 0 and $\infty$ rather than $-\infty$ and $\infty$. This allows the loss to converge to 0, providing numerical stability.   

\subsection{Training and Testing Process}

The training process follows that depicted in Figure 2. The encoding and whitening of the randomly generated bits occurs before the modulating, sampling, and shaping processes and the final transmission process is adversarial signal generation of the AMN. Before transmission, the adversarial signal is normalized so that the average symbol power is 1. PyTorch is used to implement the AMN and training process. 

\section{Results}

The communications aware attack framework just described was used to train a variety of AMNs. There are three main focuses for the results:
\begin{inparaenum}
    \item to determine the difference in communication and evasion capabilities between AMNs trained with and without FEC,
    \item to study the effects of the adaptations made on the communication aware framework when compared to prior work, and
    \item to understand the impact of changing $\gamma$ (power loss constant) and thus varying the power, and therefore impact, of the perturbation.
\end{inparaenum}
For ease of analysis, the $\alpha$ (adversarial loss constant) and $\beta$ (communication loss constant) values were fixed at 70\% and 30\% of the remaining $1-\gamma$ respectively (since the three losses are set to sum to 1). These values were set such that evasion was prioritized adequately regardless of the value of $\gamma$. Otherwise, the transmitter would not successfully evade signal classification for larger $\gamma$.

\subsection{Intelligent Perturbations with FEC}

As shown in prior works, the addition of FEC inherently improves the intended communication link during an evasion attack given its inherent ability to correct bit errors. However, this work aims to show that the developed improvements to the communications aware attack improve the performance beyond the FEC's capabilities acting alone.
To demonstrate this, Figure 3a shows the results of both the framework considering FEC during training and the framework when FEC is taken out of the training process for a $\gamma$ set to 0.1 (shown with the solid lines). The modulation scheme and FEC coding are QPSK and Hamming (7,4) respectively. As can be seen, there is improved intended communication performance given that the SNR differences between the BER curves for Hamming (7,4) and non-coded communication links are further apart than their respective theory curves in the code's operating region. For example, the SNR required to achieve a BER of $10^{-3}$ using an AMN trained with FEC has an improvement of roughly 1.5 - 2 dB over an AMN trained without. To further emphasize the improved framework's ability to inherently leverage the FEC code, this figure also shows the BER curve when the transmitted signal utilizes a Hamming (7,4) code but is perturbed using an AMN that was trained without FEC. In this case, there is still a roughly 25\% improvement in communication performance. These results show that \emph{the AMN has learned how to more intelligently craft the perturbation when FEC is present in the training process such that it limits the hit on communications performance.} 

\begin{figure*} 
    \centering
  \subfloat[Hamming (7,4)\label{1a}]{%
       \includegraphics[width=0.32\linewidth]{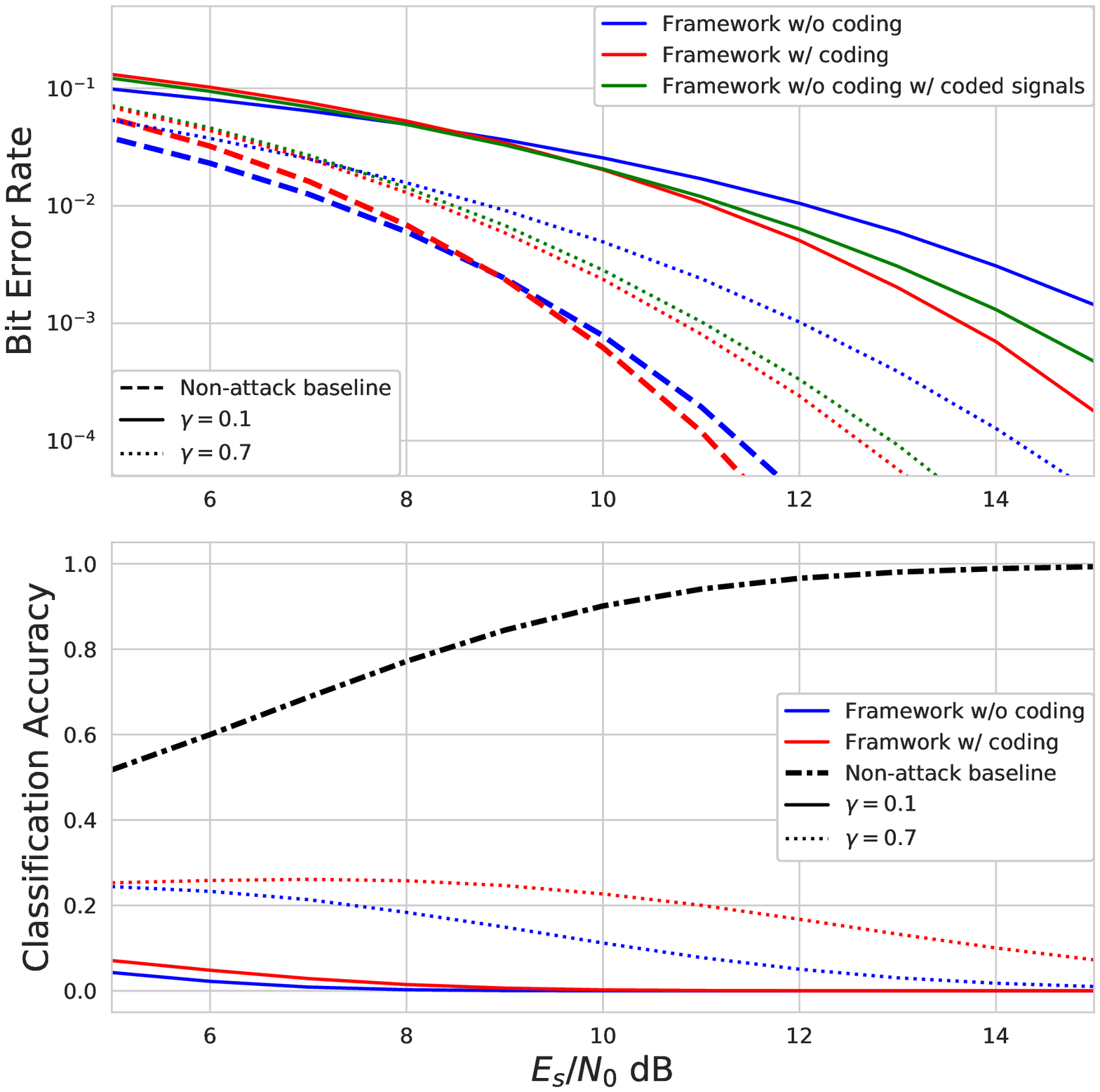}}
    \hfill
  \subfloat[BER effect held constant\label{1b}]{%
        \includegraphics[width=0.32\linewidth]{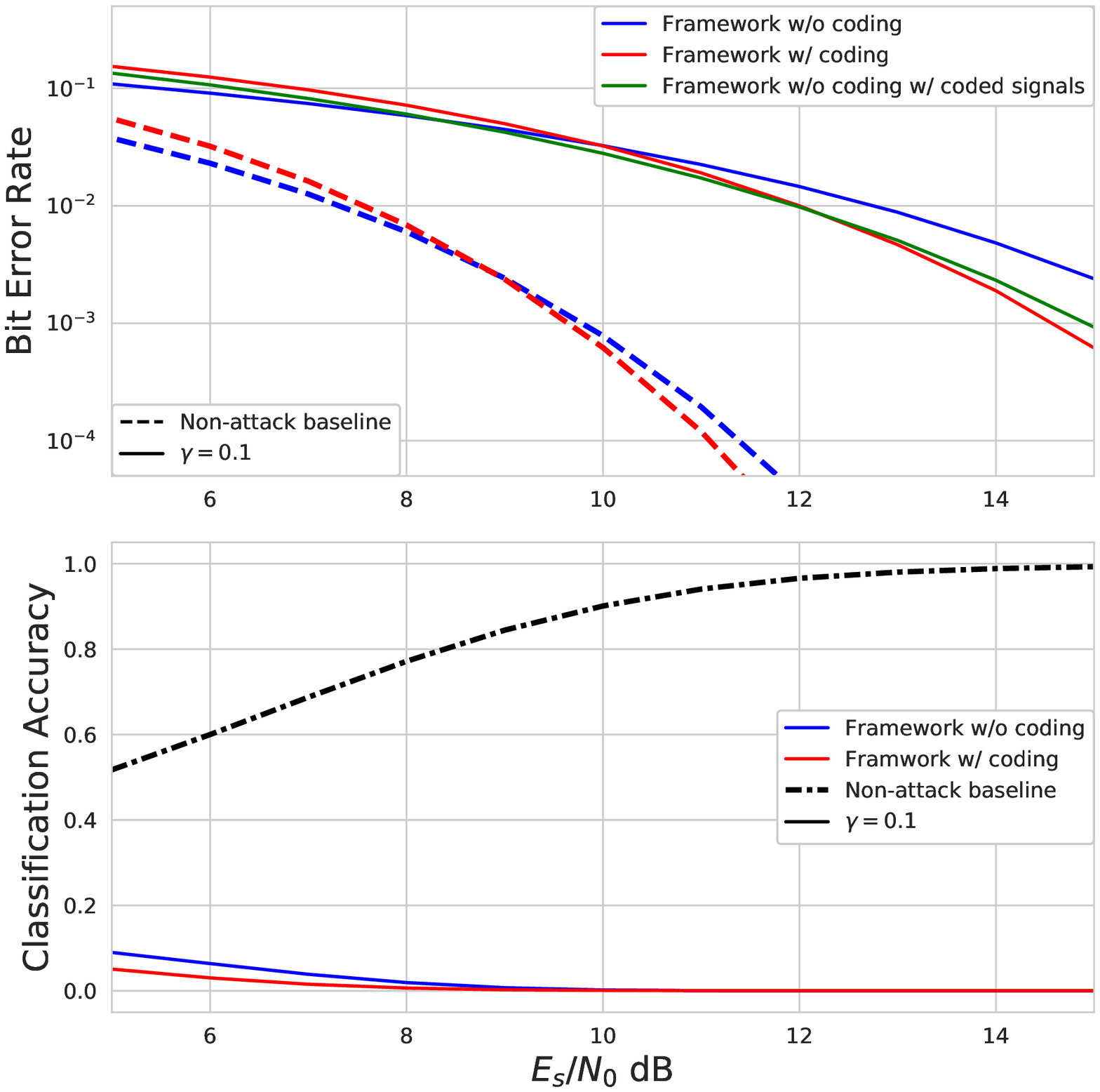}}
    \hfill
  \subfloat[Hamming (12,8)\label{1c}]{%
        \includegraphics[width=0.32\linewidth]{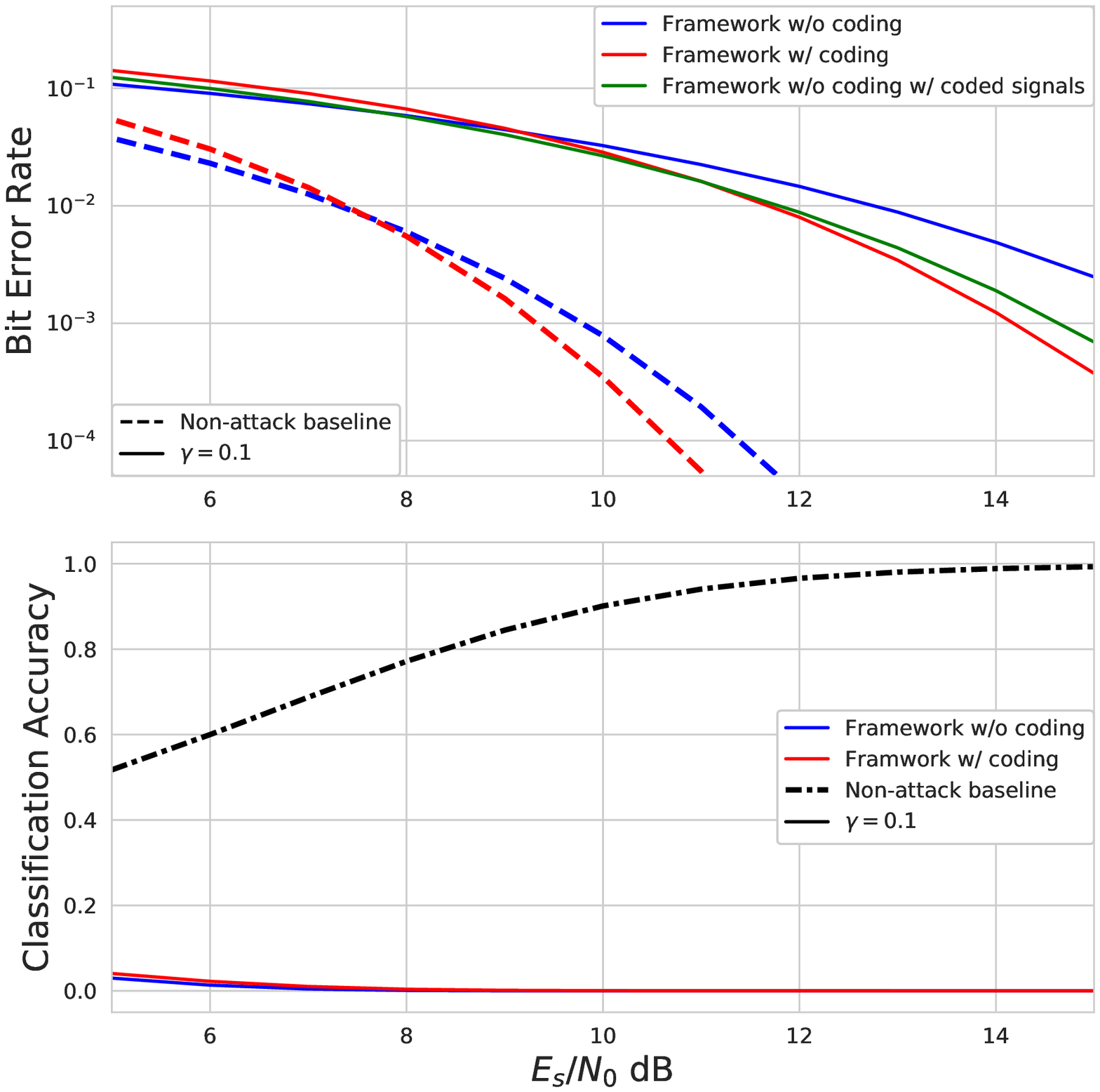}
        }
    \hfill
  \caption{Intended communications link BER and eavesdropper classification accuracy for different values of SNR ($E_{s}/N_{0}$) given a QPSK modulated signal. The baseline theoretical curves of BER are shown given both FEC coding and non-coding as well as the baseline classification accuracy of the eavesdropper with no perturbation applied. (a) performance when using Hamming (7,4) and $\gamma$ values of 0.1 and 0.7. (b) performance for Hamming (7,4) and a $\gamma$ value of 0.1 when the communication impact between coding and non-coding is more equal. (c) performance when using Hamming (12,8) and a $\gamma$ value of 0.1. In each case the addition of FEC in the training process improved communication with respect to BER with little to no degradation in the reduction of eavesdropper performance.}
  \label{fig1} 
\vspace{-1em}
\end{figure*}


While Figure 3a shows that there is a noticeable improvement in communications performance, for this case there is also an increase in the eavesdropper's classification success. However, this is a very small improvement compared to the communication improvement. Additionally, the classification accuracy of the eavesdropper is equal between the two implementations for approximately half of the SNR range.
To show this trade off further, Figure 3b demonstrates an observed case where the impact to communications between the AMN trained with FEC and that without is more equal, as indicated by the green and red lines being closer together. In other words, transmitting encoded signals using the AMN trained with FEC would offer the same BER as transmitting encoded signals using the AMN trained without FEC. When this occurs, the accuracy of the eavesdropper decreases when the AMN is trained with FEC. Therefore, if the desire is to improve the evasion performance, \emph{the transmitter trained with FEC can decrease the success of the eavesdropper's AMC if maintaining the same level of communication reliability as a transmitter trained without FEC instead of improving the reliability.} 

Figure 3c shows similar trends for a Hamming (12,8) applied to a QPSK signal. Given these results, it is shown that \emph{the AMN is able to learn to use the coding without any architecture changes} since it learns the FEC implicitly and doesn't rely on knowledge of the specific coding.

As discussed in the previous section, the $\gamma$ value represents the weighting of the loss function that controls the power of the perturbation relative to the signal. To show the impact of $\gamma$, Figure 3a also shows the results when $\gamma$ is increased from 0.1 to 0.7 (shown with dotted lines). This higher $\gamma$ value means the training process prioritizes the goal of intended communications over evasion. This effect can be seen in the figure as improved BER performance at the cost of increased classification accuracy. The improvement when training using FEC is most significant when the evasion is prioritized since the FEC-enabled AMN is more efficient with the limited communication and power losses.

Figure 4 shows the BER of the intended communications link and the classification accuracy of the eavesdropper for a variety of $\gamma$ values assuming a 16-QAM signal held at a constant SNR of 12dB. As can be observed, as $\gamma$ increases, the perturbation power loss is more prioritized which leads to the perturbed signal being closer to the original signal. As this occurs, the BER decreases but the accuracy increases as would be expected, and as is seen across all modulation schemes and FEC codes tested, ensuring generality across configurations.
\begin{figure}[t]
    \includegraphics[width=1.0\linewidth, center]{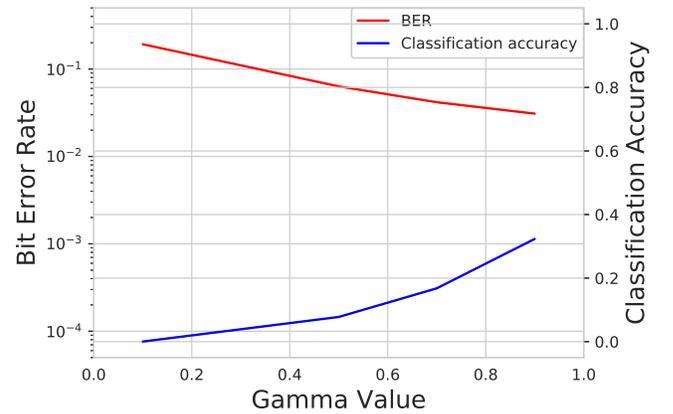} 
    \caption{Intended communications link BER and eavesdropper classification accuracy given a transmitted 16-QAM signal with SNR=12dB for different weightings of the power loss function during the communications aware attack framework's training process (represented by $\gamma$).}
    \label{g1Results}
\vspace{-1em}
\end{figure}


\subsection{Spectral Improvement}

\begin{figure} 
\centering
  \subfloat[Power Spectral Density\label{1a}]{%
       \includegraphics[width=1.0\linewidth]{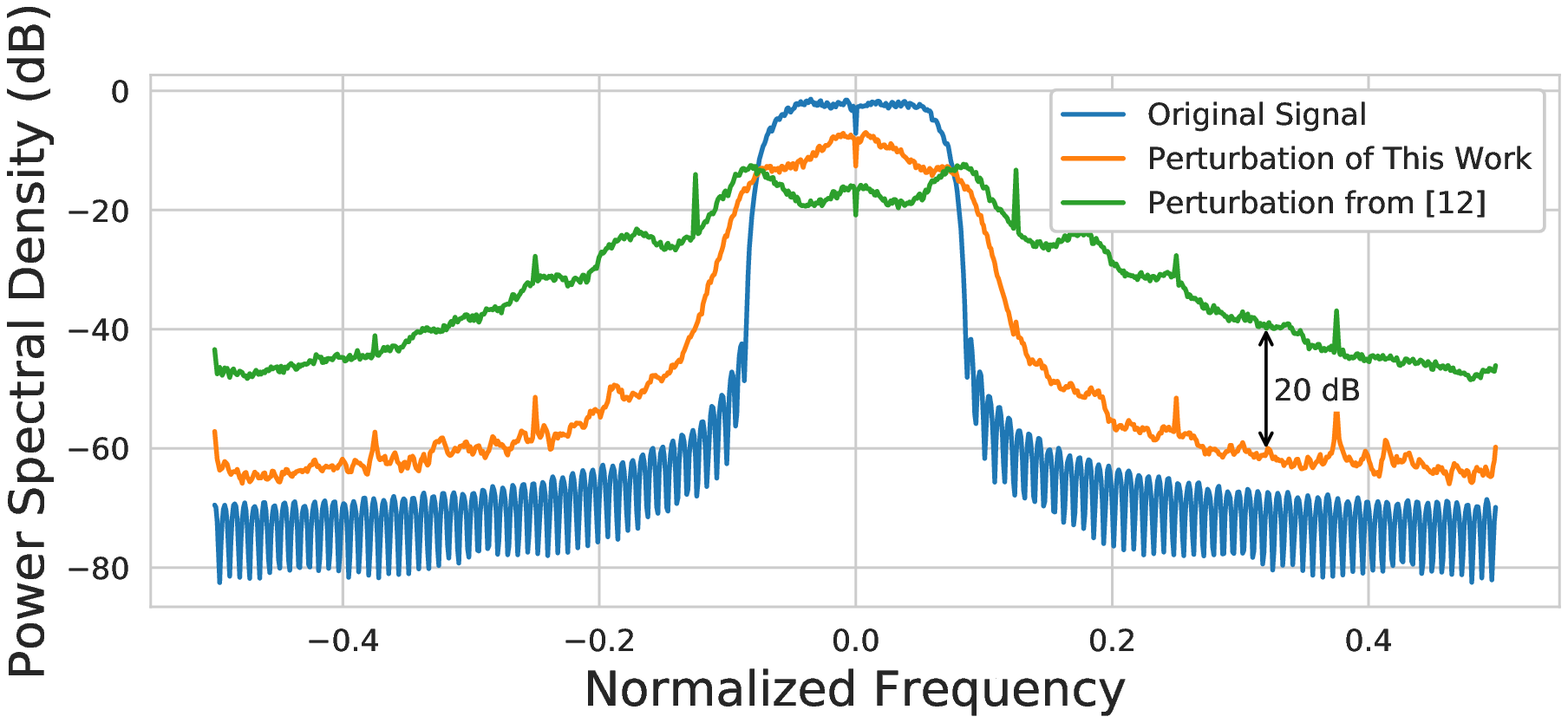}}
    \\
\vspace{-1em}
  \subfloat[Time-Domain Representation\label{1b}]{%
        \includegraphics[width=1.0\linewidth]{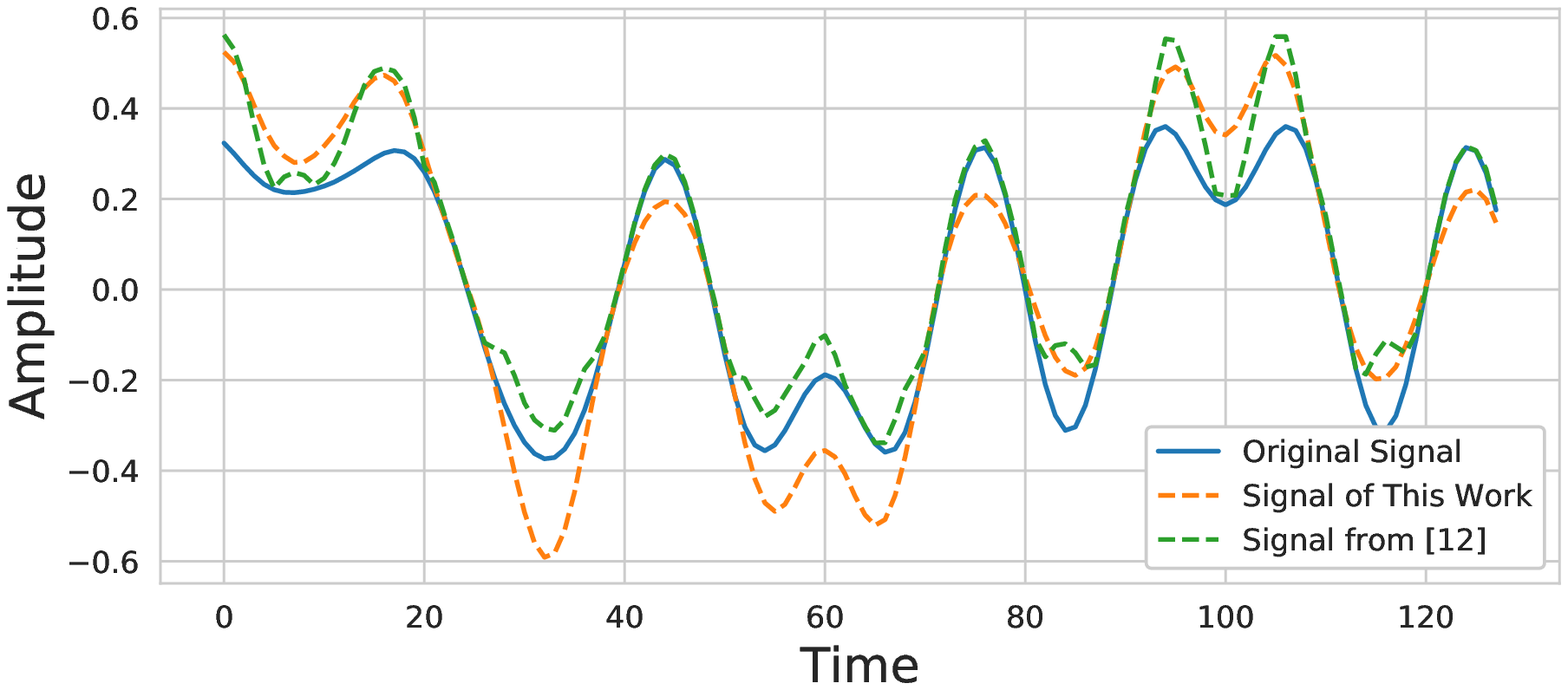}}
  \caption{The (a) spectral shape and (b) time-domain representation of a transmitted QPSK signal with and without perturbation. The improved communications aware framework developed in this work reduces the out-of-band effects caused by the perturbation over the prior work.}
  \label{fig1} 
  \vspace{-1em}
\end{figure}


The communications aware attack framework developed in the previous work tended to spread the perturbation out-of-band of the transmitted signal to reduce impact on the communications link, given the assumption of oversampling at the eavesdropper (an intuitive result indicating correct functionality) \cite{RN2}. Figure 5a shows the spectrum usage of the signal and perturbation for both the current and previous framework in \cite{RN2}. As can be seen, this work improves on the spectral efficiency of the perturbation by moving more of the perturbations in band and reducing out-of-band emissions. While there is still perturbation outside the main band, it is less significant in power and the perturbation more closely follows the frequency structure of the original signal. These improvements are due to the framework updates of using an AAE and the power loss. Figure 5b shows a plot of the signal in the time domain for the imaginary channel with the original signal, the signal generated \cite{RN2}, and the signal generated with this work. The signal created using the framework developed in this work more closely resembles the original signal in structure. The adversarial signal in previous work is less smooth due to its high frequency content, especially in the minima. Additionally, the signal power given this approach is equal to the original signal while the previous work generates a signal with about an 8-10\% increase in power.  Therefore, \emph{this work presents an attack that is both more spectral efficient and more power efficient than the prior work}.

\section{Conclusion and Future Work}

This work has shown that the communications-aware attack framework, trained with signals utilizing FEC, can inherently learn to leverage the added data redundancy to generate more intelligent perturbations that have less impact on the intended communication link while not impacting evasion performance against an eavesdropper. To achieve this, modifications to the framework were developed that allow for improved feedback through the training loss functions that more directly represent the impact of FEC on the intended communications link. Performance analysis shows that for the operating region of the FEC code, the improved framework developed in this work was able to better evade the eavesdropper for a given intended communications link bit-error rate over a system not utilizing FEC. The results of this work demonstrate that the improved performance is not just due to the inherent benefits of using FEC on the communications link, but also due to the framework intelligently learning to manipulate the transmitted signal based upon the capabilities of the given FEC code.

In addition to the enhanced attack performance, the improved framework developed in this work provides for perturbed signals that better hold their original spectral shape than 
what was seen in the prior work \cite{RN2}. A limitation of the prior work was the assumption that the eavesdropper oversampled the received signal allowing out-of-band perturbations effects. This improved framework therefore allows for both relaxed eavesdropper assumptions and more efficient bandwidth utilization of the perturbed transmitted signal.


While this work focused on the inherent benefits of using FEC in the framework, the results herein could be further improved through utilizing explicit knowledge of the FEC coding scheme during the training process. More specifically, targeted updates to the loss functions and AMN architecture design decisions (such as setting convolution kernel sizes and stride lengths relative to the FEC code) that correspond to the specifics of the FEC code used could further enhance the performance of the framework. Additionally, future work should work to further improve the spectral characteristics of the perturbed signal through the incorporation of loss metrics that aim to keep the transmitted signal within its predefined spectral mask. This would have the added benefit of improved performance against eavesdroppers with intelligent filtering processes aimed at removing out-of-band perturbations. Other targets for future work include incorporation of knowledge of the channel propagation effects between the intended receiver and/or eavesdropper, as was done in \cite{RN26}, as well as the impacts of non-block code FEC schemes on the framework.

\bibliographystyle{IEEEtran}
\bibliography{man}
\vspace{12pt}

\end{document}